\documentclass[twocolumn,aps,pre,amsmath,showpacs]{revtex4-1}
\usepackage{graphicx,dcolumn,bm,amssymb,amsmath,ulem,indentfirst,amsthm}
\usepackage[toc,page]{appendix}
\usepackage{color}

\theoremstyle{plain}
\def\be{\begin{equation}}
\def\ee{\end{equation}}

\newtheorem*{theorem*}{Theorem}

\begin{document}

\author{Bingyu Cui$^{1,2}$, Rico Milkus$^{1}$, Alessio Zaccone$^{1,3}$}
\affiliation{${}^1$Statistical Physics Group, Department of Chemical
Engineering and Biotechnology, University of Cambridge, New Museums Site, CB2
3RA Cambridge, U.K.}
\affiliation{${}^2$Department of Applied Mathematics and Theoretical Physics,
University of Cambridge,
Wilberforce Road, Cambridge CB3 0WA, U.K.}
\affiliation{${}^3$Cavendish Laboratory, University of Cambridge, JJ Thomson
Avenue, CB3 0HE Cambridge,
U.K.}
\begin{abstract}
We compute the dielectric response of glasses starting from a microscopic system-bath Hamiltonian of the Zwanzig-Caldeira-Leggett type and using an ansatz from kinetic theory for the memory function in the resulting Generalized Langevin Equation.
The resulting framework requires the knowledge of the vibrational density of states (DOS) as input, that we take from numerical evaluation of a marginally-stable harmonic disordered lattice, featuring a strong
boson peak (excess of soft modes over Debye $\sim\omega_{p}^{2}$ law). The dielectric
function calculated based on this ansatz is compared with experimental data for the paradigmatic case of glycerol at
$T\lesssim T_{g}$. Good agreement is found for both the reactive (real part) of the response and for the $\alpha$-relaxation
peak in the imaginary part, with a significant improvement over earlier theoretical approaches, especially in the reactive modulus.
On the low-frequency side
of the $\alpha$-peak,  the fitting supports the presence of $\sim \omega_{p}^{4}$ modes at vanishing eigenfrequency as recently shown in [Phys. Rev. Lett. 117, 035501 (2016)]. $\alpha$-wing asymmetry and stretched-exponential behaviour are
recovered by our framework, which shows that these features are, to a large extent, caused by the soft boson-peak modes in the DOS.

\end{abstract}

\title{Direct link between boson-peak modes and dielectric $\alpha$-relaxation in
glasses}
\maketitle

\section{Introduction}
Supercooled liquids that undergo a liquid-glass transition exhibit an abrupt
and dramatic slowdown of the atomic/molecular dynamics upon approaching the
glass transition temperature $T_{g}$~\cite{Donth,Ngai,Goetze,Goetze-book}.
The $\alpha$-relaxation describes the slowest component of the time-relaxation
(or autocorrelation function) of material response, including mechanical
relaxation, relaxation of density fluctuations or of the dielectric
polarization~\cite{Richert}.
The $\alpha$-relaxation phenomenon has always been associated with the
collective and strongly cooperative motion of a large number of atoms/molecules
which rearrange in a long-range correlated way~\cite{Donth}. This process has
also been interpreted, within the energy landscape picture, as the transition
of the system from one meta-basin to another, which involves the thermally
activated jump over a large energy barrier~\cite{Zamponi,Mezard,Yoshino}.

Modern theories of dielectric response of matter~\cite{Froehlich,Boettcher} are
based on the Lorentz model~\cite{Born,Born-Wolf}, which approximates electrons
as classical particles bound harmonically to positive background charges. Upon
assuming that all oscillators move at the same natural frequency, the
relaxation function
$\epsilon(t)$ is a simple-exponential increasing function of time, while the
imaginary part $\epsilon''(\omega)$ of the complex dielectric function
$\epsilon^{*}(\omega)$, features a resonance peak given by a Lorentzian
function~\cite{Born,Born-Wolf}.

Correcting to account for the rotational Brownian motion in the case of
strongly anisotropic molecules, as in the Debye dielectric-relaxation
theory~\cite{Froehlich}, does not alter the simple-exponential relaxation.
While this may be a good approximation for gases and high-$T$ liquids, it is
not valid for glasses, as is well known since the time of
Kohlrausch~\cite{Cardona,Williams}. For supercooled liquids in general, and for
glasses in particular, the Kohlrausch stretched-exponential function $\sim
\exp[-(t/\tau)]^{\beta}$ provides a good empirical fit of the relaxation function
and of the dielectric loss~\cite{Donth,Ngai,Phillips,Richert,deGennes,Schwartz}.

Mode-coupling theory (MCT) developed by W. Goetze and co-workers, has provided
a general interpretation of the $\alpha$-peak in dielectric relaxation using a
framework where the many-body microscopic dynamics of charges is treated
statistically, in the same way as for an ensemble of classically interacting
spherical particles~\cite{Goetze-book}. The most striking success of MCT
has been the first-principles derivation of the Kohlrausch
stretched-exponential relaxation for $\alpha$-relaxation in the liquid phase.

While MCT has had tremendous success in describing supercooled liquids at
$T>T_{g}$, the situation is quite different at $T\simeq T_{g}$ or in the glass
at $T<T_{g}$. Here, although MCT provides a theoretical foundation for Kohlrausch stretched-exponential
behaviour, direct comparisons with experimental data have not been possible due
to the difficulty of calibrating various parameters in the theory.
This scenario is the most striking for the paradigmatic case of glycerol: this is
the most widely studied organic glass-former in the experimental literature, yet no microscopic theory has been used to describe the dielectric response of this material
apart from empirical models (e.g. Havriliak-Negami), which have no physics in them.

Here we take a very different approach: instead of the liquid-state approach of
MCT, we take the opposite point of view, and describe the dynamics in analogy
with a disordered low-T lattice of particles which perform harmonic
oscillations. Due to the disorder in the lattice (and in particular due to the
absence of local inversion symmetry~\cite{Milkus}), the low-frequency part of
the vibrational density of states (DOS) is dominated by an excess of soft
modes over the Debye $\omega_p^2$ law valid for crystals.

This excess of soft modes in the DOS is universally known in the literature on glassy physics and disordered
systems as the "boson peak"~\cite{Schirmacher,Tanaka,Silbert,Liu}.
In the following we use this terminology and we refer to the broad ensemble of all these excess soft modes over the Debye
$\omega_p^2$ law as the boson peak.
It is important to note that, in the sub-field of dielectric spectroscopy of glasses, the terminology "boson peak" is used
to designate an isolated peak in the THz frequency regime of the dielectric loss modulus $\epsilon''$.
In our work we will never refer to or consider this THz-frequency peak in the loss modulus, so there is no ambiguity in our terminology
and the term "boson peak" is used exclusively to designate the ensemble of excess non-Debye modes in the low-$\omega_{p}$ part of the vibrational DOS.

Famous physicists in the past have attempted to explain stretched-exponential relaxation (which is the hallmark of the $\alpha$-relaxation
in glasses) in terms of the underlying cooperative coupling of vibrational degrees of freedom~\cite{Phillips, deGennes,Schwartz,Langer,Procaccia}.
In spite of these efforts, the link between quasi-localized soft vibrational modes or boson peak modes in the DOS, and the
$\alpha$-relaxation process, has surprisingly received less attention, with important exceptions like Ref.~\cite{Harrowell}
and the macroscopic model for viscoelasticity of Ref.~\cite{Mazzone}. This is despite the fact
that both the boson peak in the vibrational DOS and the $\alpha$-relaxation process display a strong $T$-dependence near $T_{g}$
(for the $T$-dependence of the boson peak, see e.g.~\cite{Teichler,Mitrofanov}).

For the first time, we present a simple and explicit set of relations between
the dielectric relaxation functions and the DOS of disordered lattices, based on the ansatz that the microscopic Hamiltonian can be modelled using a system-bath coupling of the Zwanzig-Caldeira-Leggett type. Within this framework,
it is possible to show that soft modes in the DOS in the boson-peak region are responsible
for the observed stretched-exponential relaxation in time and for the
$\alpha$-relaxation peak in the loss modulus of glycerol at $T\lesssim T_{g}$.

\section{Generalized Langevin Equation for dielectric response within the Lorentz model}
In the following, we work within the Lorentz dielectric model of
disordered elastically bound classical charges (basically on the same general
coarse-graining level as in Ref.~\cite{Goetze}).
We will derive an average equation of motion for a tagged charge/particle which is the microscopic building block in our treatment.
We start from an effective Hamiltonian of the Zwanzig-Caldeira-Leggett type which allows one to describe the motion of a tagged particle which is coupled to a large number of harmonic oscillators (the bath), in our case representing the other particles in the glass. As is known, the Caldeira-Leggett models~\cite{Leggett} provide an effective way of accounting for long-range anharmonic interactions, and for the resulting dissipation, by keeping the analysis within the limits of linear theory and eventually leading to a Generalized Langevin Equation (GLE) where an ansatz in form of stretched exponential is chosen for the memory kernel, for the dynamics.

This GLE is our microscopic equation of motion that we then use within the Lorentz model to obtain the dielectric function. Since this last step involves summing over all the microscopic degrees of freedom of the system, the vibrational density of states (DOS) is required for a quantitative evaluation. Since the DOS cannot be derived analytically from first-principles, we will make use of the simplest meaningful DOS $\rho(\omega_{p})$, consistent with our Hamiltonian, to represent a harmonic disordered lattice, obtained by numerical diagonalization
of a model random lattice of harmonically-bound spherical particles. This lattice is obtained by driving a
dense Lennard-Jones system into a metastable glassy energy minimum with a
Monte-Carlo relaxation algorithm, and then replacing all the nearest-neighbour
pairs with harmonic springs all of the same length and spring
constant~\cite{Milkus}. Springs are then cut at random in the lattice to
generate disordered nearest-neighbour lattices with variable mean coordination $Z$, from
$Z=9$ down to the isostatic limit $Z=2d=6$. It is important to notice that this simplified model DOS
is convenient for its simplicity and to single-out generic features of glassy behaviour, but in order to obtain very accurate fittings,
more realistic simulated DOS (e.g. from molecular simulations) will be employed in future work.

\subsection{System-bath Hamiltonian for disordered dissipative lattices}
Caldeira-Leggett models~\cite{Leggett} for the effective treatment of anharmonicity leading to the emergence of dissipative behaviour in condensed matter have been applied mostly to low-temperature quantum physics problems, particularly quantum tunnelling in superconductors and in chemical reaction rate theory. We are not aware of any application of this approach to microscopic dynamics in disordered solids such as glasses. We do this with the aim of deriving a suitable equation of motion for a tagged molecule in a low-T glass, which takes into account the disordered environment as well as the dissipation.
We use the Zwanzig-Caldeira-Leggett (ZCL) classical version of the model, which was actually proposed already in 1973 by Zwanzig~\cite{Zwanzig}, since it is particularly convenient in order to build a connection with the DOS of a disordered solid.

The general form of the Caldeira-Leggett Hamiltonian for the classical dynamics of a tagged particle that is coupled to a large number of harmonic oscillators (which, in our case, effectively represent the dynamics of all other particles in the glassy system) is given by three terms~\cite{Zwanzig}:
\begin{equation}
H=H_S+H_B+H_I
\end{equation}
where $H_S=p^2/2m+V(q)$ is the Hamiltonian of the tagged particle, $H_B=\sum_{\alpha=1}^N(\frac{p^2_{\alpha}}{2m_{\alpha}}+\frac{1}{2}m_{\alpha}\omega^2_{\alpha}x^2_{\alpha})$ is the Hamiltonian of the bath of harmonic oscillators that are coupled to the tagged particle, $H_I=\sum_{\alpha}^NH_{\alpha}(q,x_{\alpha},\omega_{\alpha},\dot{q},\ddot{q},...,\dot{x}_{\alpha},\ddot{x}_{\alpha},...)$ is the coupling term between the tagged atom and the bath. Note we also assume that there is no interaction between bath oscillators. Under these definitions, the equations of motion are:
\begin{equation}
m\ddot{q}=-V'(q)-\sum_{\alpha}\frac{\partial H_{\alpha}}{\partial q}
\end{equation}
for the tagged particle, and
\begin{equation}
m_{\alpha}\ddot{x}_{\alpha}+m_{\alpha}\omega_{\alpha}^2x_{\alpha}=-\frac{\partial H_{\alpha}}{x_{\alpha}}.
\end{equation}
for an oscillator that belongs to the bath.

For the ZCL model, the Hamiltonian is given by
\begin{equation}
H=\frac{p^2}{2m}+V(q)+\frac{1}{2}\sum_{\alpha=1}^N\left[\frac{p_{\alpha}^2}{m_{\alpha}}+m_{\alpha}\omega_{\alpha}^2\left(x_{\alpha}
-\frac{F_{\alpha}(q)}{m_{\alpha}\omega_{\alpha}^2}\right)^{2}\right].
\end{equation}

\subsection{Generalized Langevin Equation (GLE) of molecular motion in low-T glasses}
In the ZCL Hamiltonian, the coupling function is taken to be linear $F_{\alpha}(q)=c_{\alpha}q$, where $c_{\alpha}$ is known as the strength of coupling between the tagged atom and the $\alpha$-th bath oscillator. The bilinear coupling assumption can be related to a small-oscillation assumption in the same spirit as the harmonic approximation.
Clearly, this choice leads to a second-order inhomogeneous differential equation for the position of the $\alpha$-th oscillator of the bath. This solution can then be replaced into the equation of motion for the tagged particle, which leads to the following GLE
\begin{equation}
m\ddot{q}=-V'(q)-\int_{-\infty}^t \nu(t')\frac{dq}{dt'}dt' + F_{p}(t).
\end{equation}
where the non-Markovian friction or memory kernel $\nu(t)$ is given by:
\begin{equation}
\nu(t)=\sum_{\alpha}\frac{c_{\alpha}^{2}}{\omega_{\alpha}^{2}}\cos\omega_{\alpha}t.
\end{equation}

The thermal noise term $F_{p}$ is instead given by the initial positions and momenta of the bath oscillators. For example, if the initial conditions for the bath oscillators are taken to be Boltzmann-distributed $\sim \exp(-H_{B}/k_{B}T)$, the noise satisfies a coloured fluctuation-dissipation theorem, $\langle F_{p}(t)F_{p}(t') \rangle= k_{B}T\nu(t-t')$. In our case of a glass far below $T_{g}$, the system is certainly not thermalized at $t=0$, so that the fluctuation-dissipation theorem is not expected to hold in this form (nor in general). Since we are interested in the low-temperature behaviour below $T_{g}$, we will make the assumption that the noise is low, and set $F_{p}=0$ above. This is consistent with frozen-in molecular positions~\cite{Hubbard}, in the absence of an external driving force.

\subsection{The memory function for the microscopic friction coefficient}
The ZCL Hamiltonian does not provide any prescription to the form of the memory function $\nu(t)$, which can take any form depending on the values of the coefficients $c_{\alpha}$~\cite{Zwanzig}. This is evident from looking at Eq.(6).
Hence, a shortcoming of CL-type models, including ZCL, is that the functional form of $\nu(t)$ cannot be derived a priori for a given system, because, while the DOS is certainly an easily accessible quantity from simulations of a physical system, the spectrum of coupling constants $\{c_{\alpha}\}$ is basically a phenomenological parameter.

However, for a supercooled liquid, the time-dependent friction, which is dominated by slow collective dynamics, has been famously derived within kinetic theory (Boltzmann equation) using a mode-coupling type approximation by Sjoegren and Sjoelander ~\cite{Sjoegren} (see also Ref.\cite{Bagchi}), and is given by the following elegant expression:
\begin{equation}
\nu(t)=\frac{\rho k_{B}T}{6\pi^2 m}\int_{0}^{\infty}dk k^{4} F_{s}(k,t)[c(k)]^{2} F(k,t)
\end{equation}
where $c(q)$ is the direct correlation function of liquid-state theory, $F_{s}(q,t)$ is the self-part of the intermediate scattering function and $F(k,t)$ is the intermediate scattering function~\cite{Sjoegren}. All of these quantities are functions of the wave-vector $k$. Clearly, the integral over $k$ leaves a time-dependence of
$\nu(t)$ which is controlled by the product $F_{s}(k,t)S(k,t)$. For a chemically homogeneous system, $F_{s}(k,t)S(k,t)\sim F(k,t)^{2}$, especially in the long-time regime.
From theory and simulations, we know that in supercooled liquids $F(k,t)\sim \exp(-t/\tau)^\xi$, with values of the stretching exponent that are typically around $\xi=0.6$~\cite{Hansen}.
In turn, this gives $\nu(t)\sim \exp(-t/\tau)^b$ with $b\approx 0.3$.
Hence, in our fitting of experimental data we will take
\begin{equation}
\nu(t)=\nu_0e^{-(t/\tau)^b},
\end{equation}
where $\tau$ is a characteristic time-scale and $\nu_{0}$ is a constant pre-factor.

\subsection{Final form of the GLE}
Further, by assuming no noise in the low-T limit, we can generalize the formal GLE for the equation of motion of an isolated tagged particle to the case of particles on a 3D nearest-neighbour lattice, where the tagged particle coordinate $q$ is replaced by the position vector $\uline{r}_{i}$ on the lattice, and the local conservative force-field $-V'(q)$ is provided by the nearest-neighbours interactions, taken here to be harmonic as a good approximation in the low-T limit, such that $-V'(q) \Rightarrow -\uline{\uline{H}}_{ij}\tilde{\uline{r}}_{j}$, where summation over the repeated index $j$ is implied. Here,
$\uline{\uline{H}}_{ij}=\partial V/\partial \uline{r}_{i}\partial\uline{r}_{j}$, is the Hessian matrix.
Furthermore, in the dielectric response problem, we also have an additional external force given by the electric field acting on the particle charge (or partial charge) $q_{e}$.
Thus we get the following GLE-type equation of motion valid at low T:
\begin{equation}
m\ddot{ \uline{r}}_{i}+\int_{-\infty}^t \nu_0e^{-[(t-t')/\tau]^b}\dot{\uline{r}}_{i}dt'+\uline{\uline{H}}_{ij}\uline{r}_{j}=q_{e}\uline{E}.
\end{equation}
where $q_{e}$ is now the partial charge on the tagged particle, and $\uline{E}$ is the externally applied electric field.

This equation of motion describes the dynamics of an average particle harmonically bound locally to its nearest-neighbours, and non-locally coupled to many other particles, due to the bilinear coupling term of the ZCL Hamiltonian. This non-local coupling represents an effective way of accounting for the effect of long-range anharmonicity in real materials, and is the root cause for dissipation and for the time-dependent friction in the memory integral.
Equation (9) will be used below within the Lorentz model of dielectrics to obtain the dielectric function.

\subsection{Vibrational DOS and its T-dependence}
As anticipated above, in the Lorentz dielectric model the displacement of all particles in the applied oscillating electric field has to be
evaluated to obtain the polarization. This require a sum over all degrees of freedom of all particles, which can be done by using the vibrational density of states and
integrating over the eigenfrequency, as will shown later.
The DOS that we will use is obtained from numerical diagonalization of the simulated network described above, and is
expressed in terms of dimensionless eigenfrequencies $\omega_{p}$.
Generally, the eigenfrequency $\omega_p$ obtained from numerics and the
eigenfrequency $\omega_p^{\prime}$ of the \textit{real}
experimental systems are related via
$\omega_p^{\prime}\approx\sqrt{\kappa/m}\omega_p$ where $m$ is the effective mass of the
charged particle and $\kappa$ the spring constant, under the condition that
$\int^{\omega^{\prime}_D}_0\rho^{\prime} (\omega^{\prime}_p)d\omega^{\prime}_p=
\int^{\omega_D}_0\rho(\omega_p)d\omega_p$.
We use the constant $C=\sqrt{\kappa/m}$ as a fitting parameter. $\kappa$ and
$m$ are both equal to unity in the numerical simulation of the DOS, whereas
their values are of course different for different experimental systems (in the
case of dielectric response, $m$ is to understood as an effective mass).

Also, the DOS obtained from diagonalization of the model random networks,
depends on the average coordination number $Z$. For example, the boson peak
frequency drifts towards lower values of $\omega_{p}$ according to the scaling
$\omega_{p}^{BP}\sim (Z-6)$. Hence, $Z$ is the crucial control parameter of the
relaxation process which in a real molecular glass changes with $T$. Therefore,
in order to use our numerical DOS data in the evaluation of the dielectric
function, we need to find a physically meaningful relation between $Z$ and $T$
at the glass transition. Within this picture, $Z$ represents the effective
number of intermolecular contacts, which increases the number of positive charges
to which a negative charge is bound in the material.

In all experimental systems which measure the $T$-dependent material response,
the temperature is varied at constant pressure, which implies that thermal
expansion is important. Following previous work, we thus employ thermal
expansion ideas~\cite{Zaccone2013} to relate $Z$ and $T$. Upon introducing the
thermal expansion coefficient $\alpha_T=\frac{1}{V}(\partial{V}/\partial{T})$
and replace the total volume $V$ of the sample via the volume fraction $\phi=v
N/V$ occupied by the molecules ($v$ is the volume of one molecule), upon
integration we obtain $\ln{(1/\phi)}=\alpha_T T + const$. Approximating $Z \sim
\phi$ locally, we get $Z=Z_{0} e^{-\alpha_T T}$. Imposing that  $Z_{0}=12$, as
for FCC crystals at $T=0$ in accordance with Nernst principle, we finally get,
for glycerol, $Z\approx 6.02$ when $T=184K$. This is very close to the reported
$T_{g}$ for this material~\cite{Lunkenheimer}.

\begin{figure}
\begin{center}
\includegraphics[height=5.3cm,width=8.5cm]{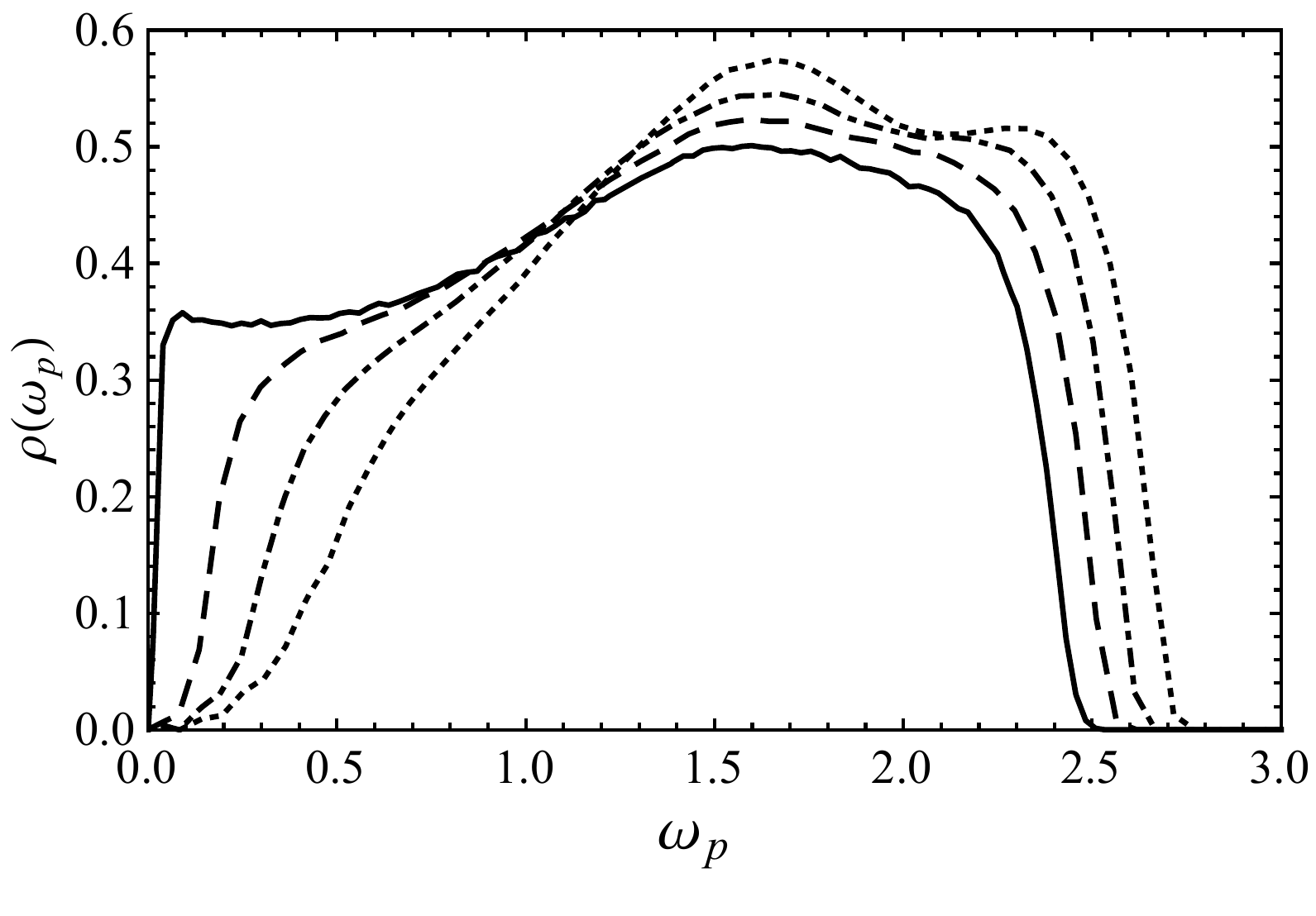}
\caption{Density of states (DOS) with respect to eigenfrequency $\omega_{p}$ at $Z=6.1$ (solid line), i.e. close to the marginal stability limit $Z=6$ that we identify here as the solid-liquid (glass) transition; plots of the DOS at $Z=7, Z=8, Z=9$ are also shown, and are marked as dashed, dot dashed and dotted lines, respectively.}
\end{center}
\end{figure}

\begin{figure}
\begin{center}
\includegraphics[height=5.3cm,width=8.5cm]{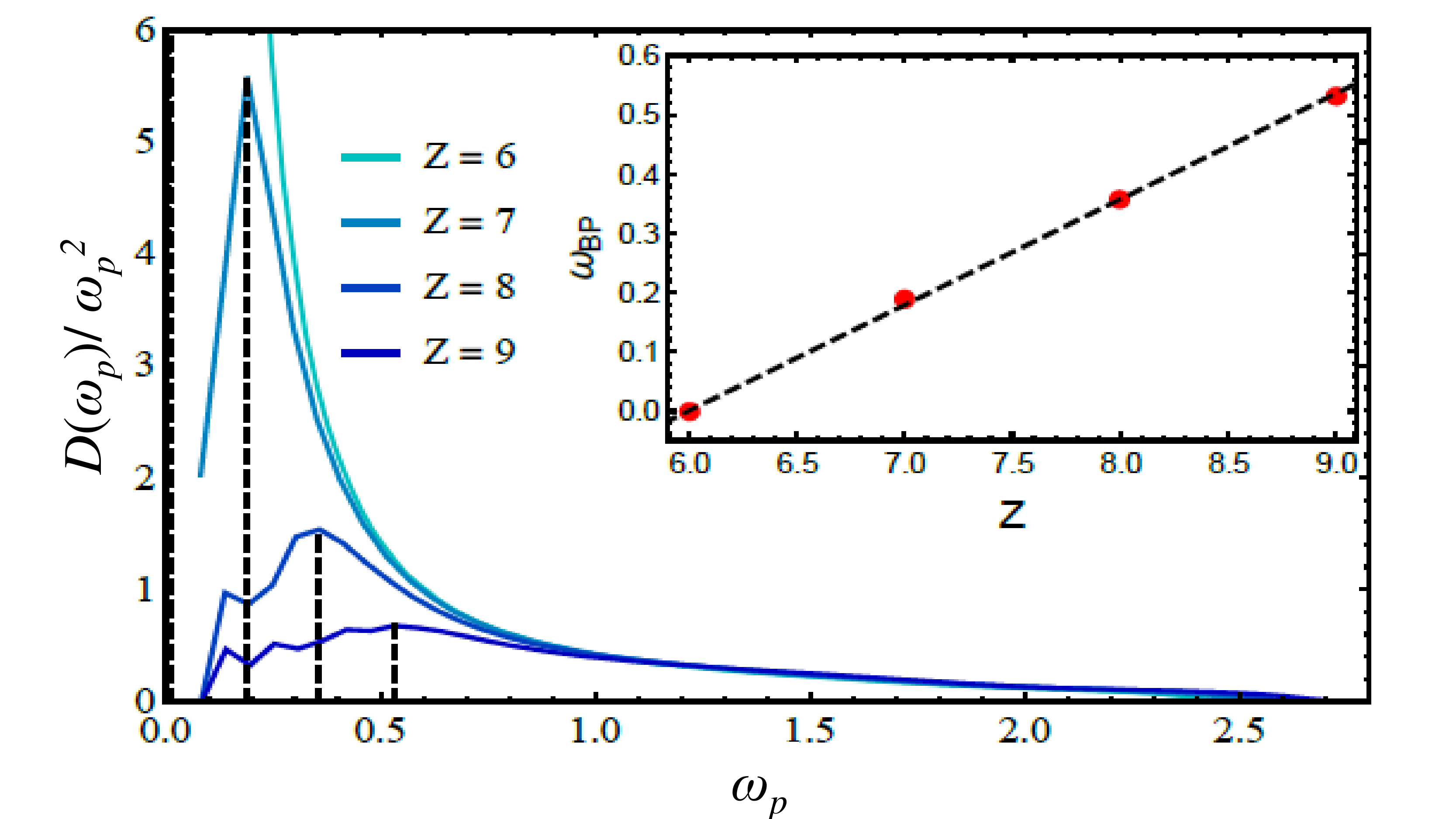}
\caption{The DOS normalized by Debye's $\omega_p^2$ law, for (from bottom to top): $Z=9, Z=8, Z=7, Z=6$, which gives evidence of the boson peak at low $\omega_p$. The eigenfrequency of boson peak scales as $\omega_p^{\textit{BP}}\sim(Z-6)$ as know from work for disordered systems with cental-force interactions~\cite{OHern,Silbert,Milkus}.}
\end{center}
\end{figure}

It is seen in Fig. 1 and in Fig. 2 that for the case $Z=6.1$, i.e. very close to the solid-liquid (glass) transition that occurs at $Z=6$, a strong and broad boson peak is present
in the DOS. Upon increasing $Z$ towards higher values the boson peak is still present but its amplitude decreases markedly upon increasing $Z$.
At $Z=6.1$, the continuum Debye regime $\sim \omega_{p}^{2}$ is not visible or
absent, whereas a very small gap between
$\omega_{p}=0$ and the lowest eigenfrequency exists. Hence, under conditions close to the glass transition where the system loses its shear rigidity, the
vibrational spectrum is dominated by a large and broad excess of soft modes with respect to Debye $\sim\omega_{p}^{2}$ law at low frequency.

\subsection{Dielectric response as a function of the vibrational DOS}
In order to determine the dependence of the polarization and of the dielectric function on the frequency of the field, we have to describe the displacement $\uline{r}$ of each molecule from its own equilibrium position under the applied field $\uline{E}$. Upon treating the dynamics classically, the equation of motion for a charge $i$ under the forces coming from interactions with other charges and from the applied electric filed, is given by the GLE Eq.(9) derived above.

The Hessian $\uline{\uline{H}}_{ij}=\partial U/\partial \uline{r}_{i}\partial\uline{r}_{j}$, where $U$ is the total potential energy of the system, represents the restoring attractive interactions from oppositely-charged nearest-neighbour charges, that tend to bring the charge $i$ back to the rest position that $i$ had at zero-field.
To solve this equation, the first step is to take the Fourier transform: $\uline{r}_{i}(t)\rightarrow \tilde{\uline{r}}_i (\omega)$, resulting in the equation:
\begin{align}
-m\omega^2\tilde{\uline{r}}_{i}+i \omega \tilde{\nu}(\omega)\tilde{\uline{r}}_{i}+\uline{\uline{H}}_{ij}\tilde{\uline{r}}_{j}&=q_{e}\tilde{\uline{E}},
\end{align}
where the tilde is used to denote Fourier-transformed variables. Hence, $\tilde{\nu}(\omega)$ is the Fourier transform of Eq.(8).

We then implement normal-mode decomposition: $\tilde{\uline{r}}_i(\omega)=\hat{\tilde{r}}_p(\omega)\mathbf{\uline{v}}_i^p$, where the hat is used to denote the coefficient
of the projected quantity, and $\mathbf{\uline{v}}_i^p$ denotes an eigenvector of the Hessian matrix. Thus the equation of motion is rewritten as
\begin{equation}
-m\omega^2\hat{\tilde{r}}_{p} +i\omega \tilde{\nu}(\omega)\hat{\tilde{r}}_{p}+m\omega_p^2\hat{\tilde{r}}_p =q_{e}\hat{\tilde{E}}, \notag
\end{equation}
where $\omega_{p}$ denotes the $p$-th normal mode frequency. The equation is solved by 
\begin{equation}
\hat{\tilde{r}}_{p}(\omega)=-\frac{q_{e}\hat{\widetilde{E}}}{m\omega^2-i\omega \tilde{\nu}(\omega)-m\omega_p^2}. \notag
\end{equation}

Upon multiplying through by the eigenvector $\mathbf{\uline{v}}_i^p$, we go back to a vector equation for the Fourier-transformed displacement of particle $i$:
\begin{equation}
\delta\tilde{\uline{r}}_{i}(\omega) =-\frac{q_{e}}{m\omega^2-i\omega\tilde{\nu}(\omega)-m\omega_{p}^2}\widetilde{\uline{E}}(\omega).
\end{equation}

Each particle contributes to the polarization a moment $\uline{p}_{i}=q_{e}\delta\uline{r}_{i}$. In order to evaluate the macroscopic polarization, we need to add together the contributions from all microscopic degrees of freedom in the system, $\uline{P}=\sum_{i}\uline{p}_{i}$.
In order to do this analytically, we use the standard procedure of replacing the discrete sum over the total $3N$ degrees of freedom of the solid with the continuous integral over the eigenfrequencies $\omega_{p}$, $\sum_{p}^{3N}...=\sum_{\alpha=1}^{3}\sum_{i=1}^{N}...\rightarrow \int\rho(\omega_{p})...d\omega_{p}$, which gives the following sum rule in integral form for the polarization in glasses
\begin{equation}
\widetilde{\uline{P}}(\omega)=-\left[\int_0^{\omega_D} \frac{\rho(\omega_p)q_{e}^2}{m\omega^2-i\omega\tilde{\nu}(\omega)-m\omega_p^2}d\omega_{p}\right]\widetilde{\uline{E}}(\omega).
\end{equation}

Here, $\rho(\omega_{p})$ is the vibrational DOS, and $\omega_{D}$ is the cut-off Debye frequency arising from the normalization of the density of states.
The complex dielectric permittivity $\epsilon^{*}$ is defined as $\epsilon^*=1+4\pi\chi_{e}$ where $\chi_{e}$ is the dielectric susceptibility which connects polarization and electric field as~\cite{Born-Wolf}: $\uline{P}=\chi_{e} \uline{E}$.  Hence, we obtain the complex dielectric function expressed as
a frequency integral as
\begin{equation}
\epsilon^*(\omega)=1-\int_{0}^{\omega_{D}}\frac{A\rho(\omega_{p})}{\omega^2-i( \tilde{\nu}(\omega)/m)\omega-C^2\omega_p^2}
d\omega_{p}
\end{equation}
where $A$ is an arbitrary positive constant, $C=\sqrt{\kappa/m}$, and $\omega_{D}$ is the Debye cut-off frequency (i.e.
the highest eigenfrequency in the vibrational DOS spectrum). As one can easily verify, if $\rho(\omega_{p})$
were given by a Dirac delta, one would recover the standard simple-exponential
(Debye) relaxation~\cite{Born-Wolf}. Note that this approach can be extended to deal with molecules
that have stronger inner polarizability by replacing the external field $\underline{E}$ with the \textit{local} electric field $\underline{E}_{\textit{loc}}$, see e.g. Ref.~\cite{Froehlich}.
The detailed derivation is provided in Appendix A. However, we have checked that the qualitative predictions are the same with and without the Lorentz field correction.

It is important to emphasize that, in Eq.(13), low-frequency soft modes which
are present in $\rho(\omega_{p})$ necessarily play an important role also at
low applied-field frequencies $\omega$, because of the $\omega^{2}$ term in the
denominator. As we will see below, this fact in our description implies a direct
role of the boson peak on the $\alpha$-relaxation process.

\subsection{Finite-size effects and low-frequency limit in the DOS}
Since we are using a DOS obtained numerically from a system with a finite
($\sim 4000$) number of particles in simulations, it is important to correctly
take care of finite size effects in Eq.(13).
In numerical simulations, the DOS
$\rho(\omega_{p})$ is not a continuous function, but discrete and can be conveniently represented as $\rho(\omega_{p})\sim\frac{1}{3N}\sum_{p=1}^{3N}\delta(\omega_{p}-\omega_{p})$. Thus, we rewrite
Eq.(13) as a sum over a discrete distribution of $\omega_{p}$:
\begin{equation}
\epsilon^*(\omega)=1-\sum_{p}\frac{A}{\omega^2-i( \tilde{\nu}(\omega)/m)\omega-C^2\omega_{p}^2}
\end{equation}
where $A$ has absorbed the scaling constant. Since the dielectric function
is a complex quantity, we can split it into its real and imaginary parts, i.e.
$\epsilon^*(\omega)=\epsilon'(\omega)-i\epsilon''(\omega)$:
\begin{align}
\epsilon'(\omega)&=\epsilon'(\infty)+\\ \notag
&\sum_p\frac{A_1(C^2\omega_{p}^2-\omega^2+\tilde{\nu}_2(\omega)\omega/m)}{(C^2\omega^2_{p}-\omega^2
+\omega\tilde{\nu}_2(\omega)/m)^2+(\omega\tilde{\nu}_1(\omega)/m)^2},
\\
\epsilon''(\omega)&=\sum_{p}\frac{A_2(\omega\tilde{\nu}_1(\omega)/m)}{(C^2\omega^2_p-\omega^2+\omega\tilde{\nu}_2(\omega)/m)^2+(\omega \tilde{\nu}_1(\omega)/m)^2}.
\end{align}
The Markovian friction case is retrieved by simply setting $\tilde{\nu}=\tilde{\nu}_1=const$ in the above expressions.
$A_1,A_2,\epsilon'(\infty), m$ are re-scaling constants that need to be
calibrated in the comparison with experimental data.
It is important to note that the experimental data of dielectric
permittivity and dielectric loss are not necessarily given in the same units and there is, in
general, no coherence between the offsets in the plots of the $\epsilon'$ and
$\epsilon''$ curves. For this reason, the values of $A_1$ and $A_2$ do not
necessarily coincide. $\tilde{\nu}_1$ and $\tilde{\nu}_2$ are real and (minus) imaginary parts of $\tilde{\nu}(\omega)$ in Fourier space, $\tilde{\nu}(\omega)=\tilde{\nu}_1(\omega)-i\tilde{\nu}_2(\omega)$.\\

As remarked above and as observed in all numerical calculations of the DOS in the vicinity of the mechanical stability point of disordered solids, there exists
a lowest non-zero eigenfrequency $\omega_{p,min}$, and a vanishingly small gap between $\omega_{p}=0$ and $\omega_{p,min}=0.019$.
A recent study has pointed out that a scaling
$\sim \omega_{p}^{4}$, possibly related to soft anharmonic modes, exists even below the lowest Goldstone modes~\cite{Lerner}. Even though we cannot settle this issue here, since it reaches far beyond the scope and interest of this work, we have performed an asymptotic analysis of the limiting behaviour of $\rho(\omega_{p})$ at $\omega_{p}\rightarrow 0$ in the context of the dielectric response. The analysis is reported in Appendix C, and clearly shows that only asymptotic scalings $\rho(\omega_{p})\sim \omega_{p}^n$, with $n>3$ can lead to meaningful behaviour of $\epsilon''(\omega)$. This finding lends further support to this form of the DOS in the zero frequency limit, and we therefore use this scaling in the finite-size gap between $\omega_{p}=0$ and $\omega_{p,min}$ which results in a linear behaviour of the left flank of the $\alpha$-peak in $\epsilon''(\omega)$ in perfect agreement with experimental data as discussed below.\\

\section{Application to dielectric relaxation data}
We now present our theoretical fittings of state-of-the-art experimental
data~\cite{Loidl,Lunkenheimer} on glycerol at $T\approx T_{g}$ using Eq.(15)-(16), also in
comparison with the empirical best-fitting Kohlrausch stretched-exponential relaxation fitting. In
Fig. 3 we plotted the comparisons for $\epsilon'(\omega)$ at $T=184~K$, i.e.
slightly below $T_{g}$, obtained by implementing the numerical DOS of Fig.1 for
$Z=6.1$ in Eq.(15).
In this case, it is clear that our theoretical model performs significantly better than the Kohlrausch best-fitting (that is optimized for the joint the fitting of dielectric loss below).
This suggests that excess soft modes are important for the fitting of the dielectric response at the glass transition.

In Fig. 4 we present fittings of the dielectric loss, $\epsilon''(\omega)$ for the Markovian case $\tilde{\nu}=\tilde{\nu}_1=const$ in Eq.(16). In
this case, it is seen that our framework even in its Markovian-friction version,  provides a reasonably good fitting of the $\alpha$-peak
on both the left-hand and the right-hand side of the peak, and the overall quality of the fitting is comparable to the one of the Kohlrausch best empirical fitting. Our theoretical model provides the crucial connection between the salient features of the DOS near $T_{g}$ and the corresponding features of the response.
Of course, at the higher-frequency end of the $\alpha$-wing, other effects may as well be important which are not described by our model: in particular, the existence of Johari-Goldstein $\beta$-relaxation-type contributions to the loss modulus in this regime has been shown for a variety of systems~\cite{Paluch,Schneider, Dob, Mattsson, Blochowicz}.

On the left-hand ascending side of the peak, the scaling $\rho(\omega_{p})\sim\omega_{p}^{4}$ leads to the linear behaviour
$\sim \omega^{1}$, as derived in Appendix C, for the ascending part of the peak.
On the high-$\omega$ side of the peak, where the dynamics is dominated by the
soft boson-peak modes and the DOS is approximately flat as a function of $\omega_{p}$ in Fig.1, our model, reproduces, remarkably, the asymmetric $\alpha$-wing behaviour still in good agreement with the experimental data.

In Fig. 5, we present the same fitting, but now with a non-Markovian friction given by $\nu(t)=\nu_0 e^{-4 t^b}$ used in Eq.(16), with $b=0.3$ suggested by previous studies on glassy dynamics. Overall, the non-Markovian friction provides a better fitting, which suggests that memory effects in the atomic dynamics are non-negligible. However, the memory kernel due to non-Markovian friction does not appear to be essential to generate and reproduce the $\alpha$-wing asymmetry.

This comparative analysis therefore demonstrates quite clearly that while memory effects are important, the main cause for the $\alpha$-wing asymmetry is the excess of soft vibrational modes in the DOS, which is a very important outcome of our study.

\begin{figure}
\begin{center}
\includegraphics[height=5.6cm,width=8.5cm]{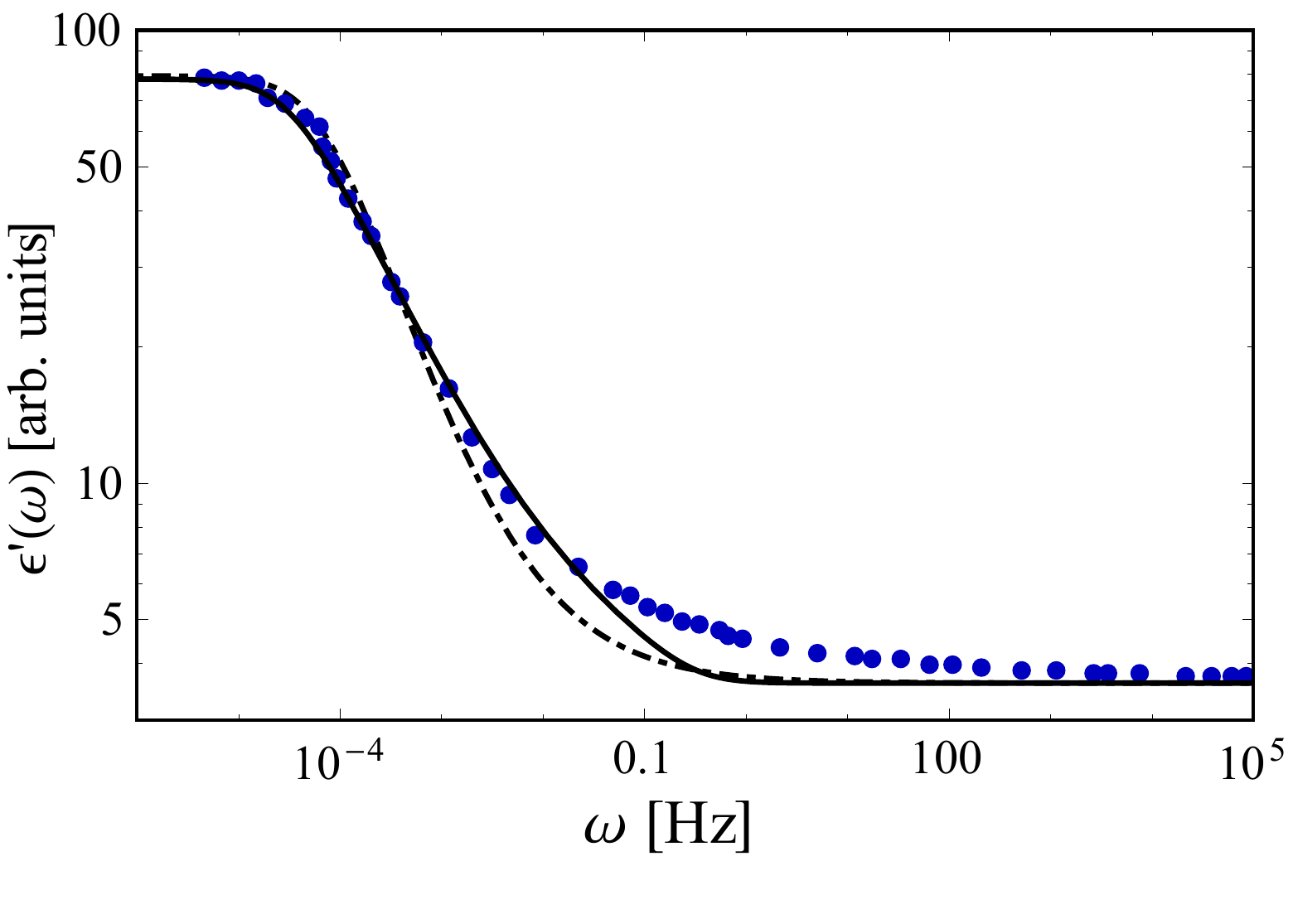}
\caption{Real part of the dielectric function as a function of the frequency of
the applied field. Symbols are experimental data of the real part of the
dielectric function of glycerol at $T=184~K$ from Ref.~\cite{Loidl}. The solid
line is our theoretical calculation for the Markovian friction case, i.e. $\nu=const$ in Eq.(15). The dot-dashed line is the real part of the Fourier transform when we consider the best-fitting (empirical) stretched-exponential function with $\beta=0.65$. We have taken $C=10, \nu/m=1620$ and $A_{1}=0.039$. For the empirical fitting with $\beta=0.65$, $\tau=6555$. Rescaling constants are used to adjust the height of the curves.}
\end{center}
\end{figure}
\begin{figure}
\begin{center}
\includegraphics[height=5.5cm,width=8.5cm]{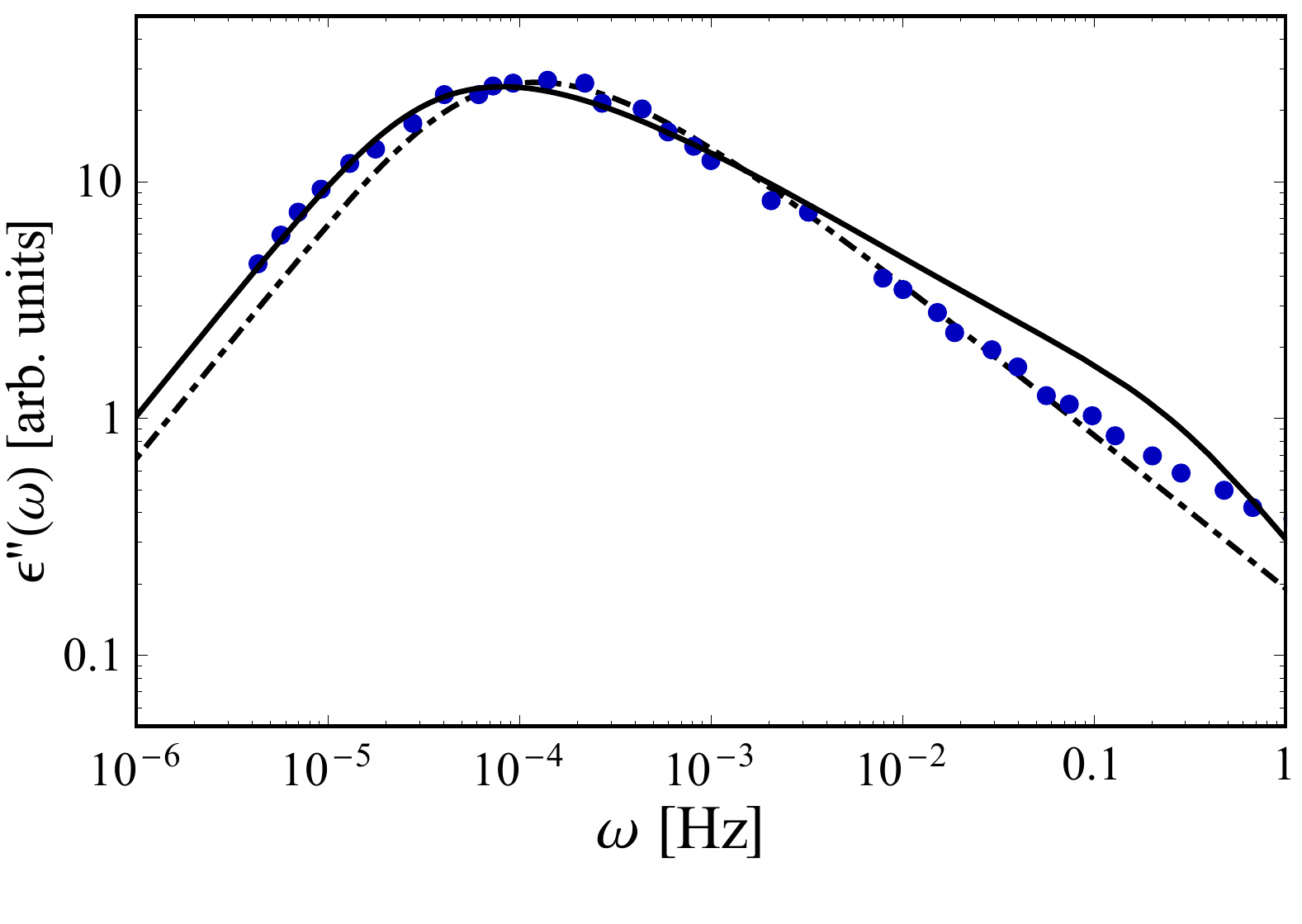}
\caption{Dielectric loss modulus as a function of the frequency of the applied
field. Symbols are experimental data of the imaginary part of dielectric
function of glycerol at $T=184K$ from Ref.~\cite{Loidl}. The solid line is our calculation for the Markovian friction case, i.e. $\nu=const$ in Eq.(16). The dot-dashed line is the imaginary part of the Fourier transform when we consider the best-fitting (empirical) stretched exponential (Kohlrausch) function with $\beta=0.65$. In our
calculation, we have taken $C=10, \nu/m=1620$ and $A_{2}=0.0437$. For the empirical fitting $\beta=0.65$, $\tau=6555$. Rescaling constants are used to adjust the height of the curves.}
\end{center}
\end{figure}

\begin{figure}
\begin{center}
\includegraphics[height=5.5cm,width=8.5cm]{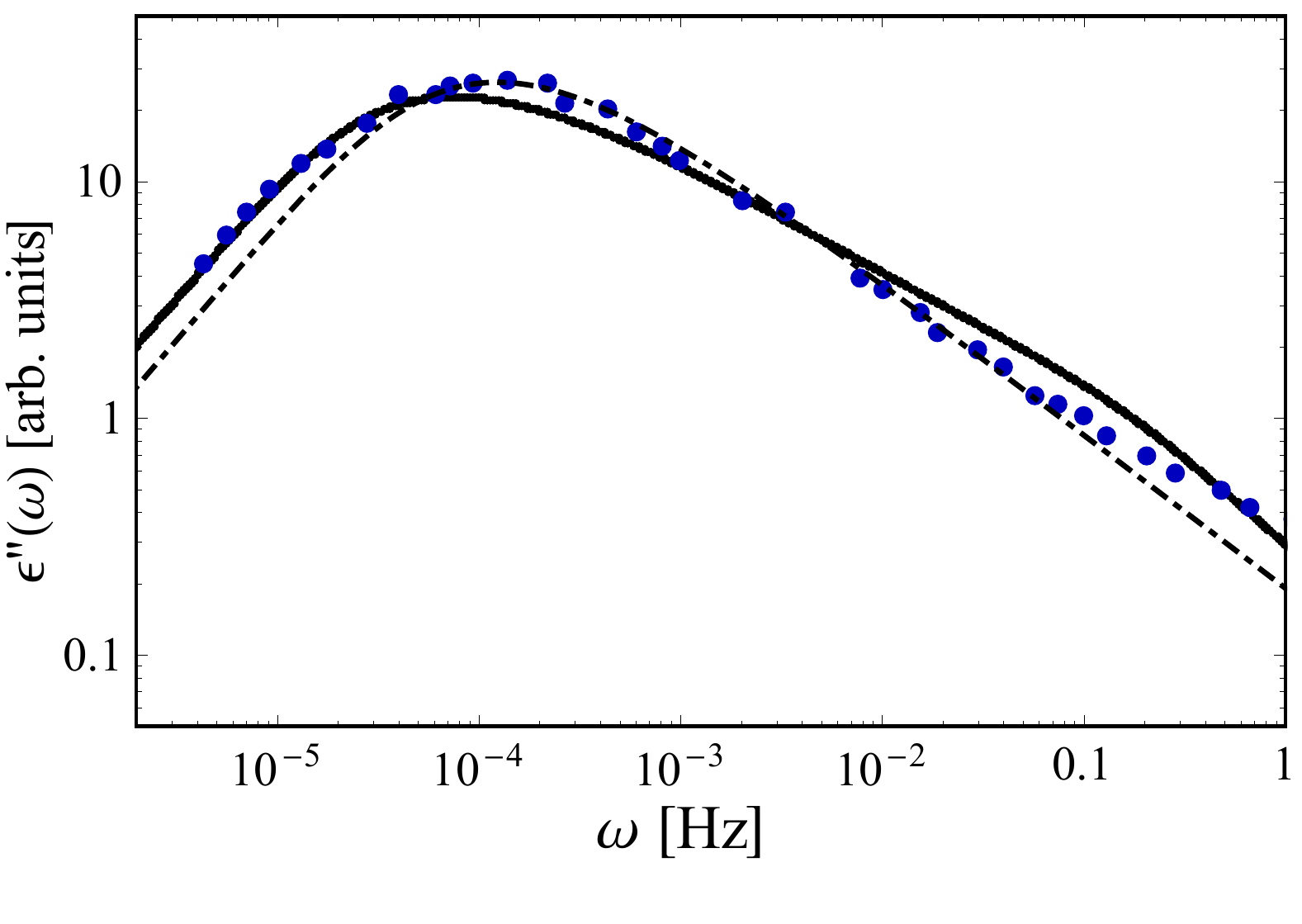}
\caption{Dielectric loss modulus as a function of the frequency of the applied
field. Symbols are experimental data of the imaginary part of dielectric
function of glycerol at $T=184K$ from Ref.~\cite{Loidl}. The solid line is the
theoretical description presented in this work for the non-Markovian friction case, i.e. $\tilde{\nu}(\omega)$ in Eq.(16) is the Fourier transform of Eq.(8). The dot-dashed line is the imaginary part of the Fourier transform when we consider the best-fitting (empirical) Kohlrausch relaxation function $\epsilon(t)\sim \exp(-t/\tau)^{\beta}$ with $\beta=0.65$. In our
calculation, we have taken $C=555$, $A_{2}=120$, $\nu_0=6\times 10^{6}$, and $b=0.3$(for the physical justification of the latter value, see Section IIC). For the empirical Kohlrausch fitting $\beta=0.65$, $\tau=6555$, as before. Rescaling constants are used to adjust the height of the curves.}
\end{center}
\end{figure}

\section{Dielectric relaxation in the time domain}
We are also interested in the dielectric response in the time domain.
In order to keep the derivation amenable to analytical treatment, we focus on the case of Markovian friction, $\nu=const$.
The time dependent
dielectric function $\epsilon(t)$ and complex dielectric function $\epsilon^*(\omega)$ are related
as:
\begin{align}
\frac{d\epsilon(t)}{dt}&=\frac{1}{2\pi}\int_{-\infty}^{\infty}(\epsilon^*(\omega)-\epsilon(\omega=\infty))e^{i\omega t}d\omega \\
\epsilon^*(\omega)&=\epsilon(\omega=\infty)-\int^{\infty}_0\frac{d\epsilon(t)}{dt}e^{-i\omega t}dt
\end{align}.

By using Eq.(13) for Markovian friction, we can write the analytical form of $\epsilon(t)$ as follows (see Appendix B for the details of the derivation):
\begin{multline}
\epsilon(t)=B+{}\\
\int_0^{\omega_D}
\frac{A\rho(\omega_p)}{2K}\left(\frac{e^{(K-\nu/2m)t}}{K-\nu/2m}+\frac{e^{-(K+\nu/2m)t}}{K+\nu/2m}\right) d\omega_p,
\end{multline}
where $K\doteq\sqrt{(C\omega_p)^2-\frac{\nu^2}{4m^2}}$, while $B$ is a re-scaling constant.
This equation is a key result: it provides a direct and quantitative relation
between the macroscopic relaxation function of the material and the DOS. As we show below, the
presence of a boson peak in $\rho(\omega_{p})$ directly causes
stretched-exponential decay in $\epsilon(t)$ via Eq.(19).

In Fig. 6, we plot predictions of Eq.(19) with the parameters calibrated in the
glycerol data fitting for the case $\nu=const$, along with the Kohlrausch function~\cite{note,Montroll},
for the relaxation in the time domain.
It is seen that our description based on soft modes is able to perfectly recover
stretched-exponential relaxation, with stretching-exponent $\beta=0.56$, over many decades in frequency. Without the
boson-peak modes in the DOS, we have checked that stretched-exponential relaxation cannot
be recovered, and the decay is simple-exponential. Hence, our Eq.(19) provides
the long-sought cause-effect relationship between soft modes and
stretched-exponential relaxation, even for the simple case of Markovian friction where the response is clearly dominated by the density of states.

\begin{figure}
\begin{center}
\includegraphics[height=5.3cm,width=8.5cm]{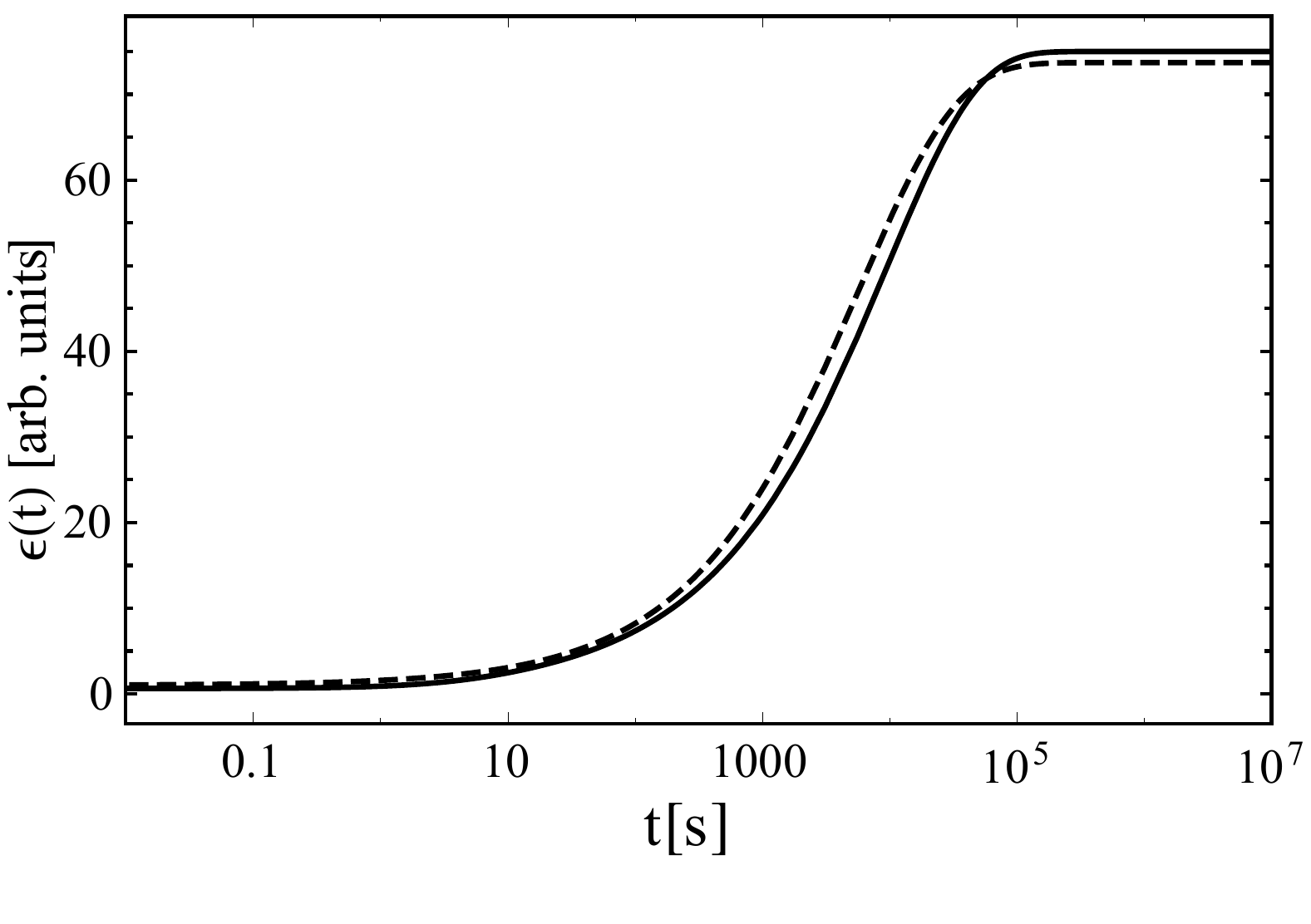}
\caption{Time-dependent dielectric response. The solid line is calculated using our Eq.(19) with physical parameters calibrated in the fitting of
Fig.3. The dashed line represents the stretched-exponential Kohlrausch function that more closely approximates our prediction,
calculated using the parameters $\beta=0.56$ and $\tau=5655$. Rescaling constants are used to adjust the height of the curves.
}
\end{center}
\end{figure}

\section{Conclusions}
We have re-considered the nature of the dielectric $\alpha$-relaxation of simple
glass-formers from the standpoint of soft modes, dissipation and lattice dynamics.
We
started from the same presumption of Ref.~\cite{Goetze} that dielectric
$\alpha$-relaxation emerges from many-body dynamics in a statistical way,
transcending the details of charge dynamics. This should especially be true
for glycerol, a paradigmatic glass-forming molecule.
Starting from a microscopic system-bath Hamiltonian, we are able to reproduce the dielectric response of glycerol
in reasonable agreement with state-of-the-art experimental data. Our physically-informed theoretical fitting compares well with empirical Kohlrausch fittings. For both the reactive modulus $\epsilon'(\omega)$ and the loss modulus $\epsilon''(\omega)$, our model provides a significantly better fitting than the best Kohlrausch fitting. For the loss modulus, our model fitting is able to reproduce the $\alpha$-wing asymmetry and although memory effects in the atomic-scale dissipative dynamics are non-negligible, the most important factor that causes the asymmetry is represented by the excess of soft modes in the DOS (boson peak). For the loss modulus $\epsilon''(\omega)$, on the low-frequency side of the $\alpha$-peak, we show that the response is controlled by the scaling $\sim \omega_{p}^{4}$ as $\omega_{p}\rightarrow 0$ in the DOS.
In the time-domain response, remarkably, our framework recovers a stretched-exponential
relaxation with $\beta=0.56$, over the entire time domain. These unprecedented results show, for the first
time, that stretched-exponential relaxation in glasses is directly caused by
the quasi-localized boson-peak excess modes contribution to the relaxation spectrum.
These results open up new opportunities to understand the crucial link between
$\alpha$-relaxation, boson peak and dynamical
heterogeneity~\cite{Chandler,Garrahan} in glasses.

\begin{acknowledgements}
Many useful discussions with R. Richert, W. Goetze, and with E. M. Terentjev are gratefully acknowledged.
\end{acknowledgements}

\begin{appendix}
\section{Lorentz local field effect on the total polarization}
In condensed matter systems, the electric field that effectively acts on a molecule locally is equal to the external field only in the limit of vanishing polarizability of the molecule~\cite{Froehlich}. This is a well-known effect whereby the field in the medium is affected (diminished) by the local alignment of the polarized molecules. The simple Lorentz cavity model works well in materials where the building blocks are not pathologically shaped or anisotropic, and is applicable to random isotropic distribution of the building blocks. Without loss of generality, we present an analysis for the case of Markovian friction $\nu=const$.
The derivation of the local field or Lorentz field can be found in many textbooks, e.g. in Refs.~\cite{Froehlich, Choy}
\begin{equation}
\underline{E}_{\textit{loc}}=\underline{E}+\frac{4\pi}{3}\underline{P}.
\end{equation}
\\

With $\underline{E}$ replaced by $\underline{E}_{\textit{loc}}$, we now write equation of motion as
\begin{equation}
m\ddot{ \uline{r}}_{i}+\nu\dot{\uline{r}}_{i}+\uline{\uline{H}}_{ij}\uline{r}_{j}=q_{e}(\underline{E}+\frac{4\pi}{3}\underline{P}).
\end{equation}
As a consequence, we instead have
\begin{equation}
\delta\tilde{\underline{r}}_{i}(\omega)=\frac{q}{m\omega^2-i\omega\nu-
\omega_{p}^2}(\tilde{\underline{E}}(\omega)+\frac{4\pi}{3}\underline{\tilde{P}}(\omega)).
\end{equation}

The total polarization is
\begin{equation}
\underline{P}=(\sum_{i} q \delta\uline{r}_{i} + \alpha \underline{E}_{\textit{loc}}),
\end{equation}
where $\alpha$ is the microscopic electronic polarizability. Combining the above relations together and summing over all contributions from all the building blocks, we obtain
\begin{align}
\epsilon(\omega)=1+4\pi\frac{\chi(\omega)}{1-\frac{4\pi}{3}\chi(\omega)},\notag\\
\chi(\omega)=q_{e}^2\int_0^{\omega_D}\frac{\rho(\omega_p)}{m\omega^2-m\omega_p^2+i\omega\nu}d\omega_p+\alpha
\end{align}
where we used $\underline{D}=\epsilon\underline{E}=\underline{E}+4\pi\underline{P}$.
We have checked that accounting for the Lorentz field and using Eq.(A5) for the fitting produces very similar results and does not alter the fitting of the dielectric relaxation data qualitatively.

\section{Derivation of Eq.(19) for the relaxation function in the time-domain}
We recall that the Fourier transform of a function $f(x)$, in this article, is defined as:
\begin{equation}
\hat{f}(\omega)=\int_{-\infty}^{\infty} f(x)e^{-i\omega x} dx
\end{equation}
while the inverse Fourier transform is defined as
\begin{equation}
f(x)=\frac{1}{2\pi}\int_{-\infty}^{\infty} f(\omega)e^{i\omega x} d\omega.
\end{equation}
From Eq.(17) in the main article, we firstly need to find the time derivative of $\epsilon(t)$:
\begin{equation}
\frac{d\epsilon(t)}{dt}=-\frac{1}{2\pi}\int_{0}^{\infty}\int_0^{\omega_D} \frac{A\rho(\omega_p)e^{i\omega t}}{\omega^2-(C\omega_p)^2-i\omega\nu/m} d\omega_p d\omega.
\end{equation}
We can change the order of integration, which gives:
\[
\int_{0}^{\omega_D}A\rho(\omega_p)\int_{0}^{\infty}-\frac{1}{2\pi}\frac{e^{i\omega t}}{\omega^2-(C\omega_p)^2-i\omega\nu/m} d\omega d\omega_p.
\]
Note that, for the inner integration, i.e., $\int_{0}^{\infty}-\frac{1}{2\pi}\frac{e^{i\omega t}}{\omega^2-(C\omega_p)^2-i\omega\nu/m} d\omega$, we could make an analytic continuation of $\omega$ to the complex plane and use contour integration to evaluate the complex integral. However, we can achieve the same result via a simpler route just using the Fourier inversion theorem~\cite{Folland} that we recall below.

\begin{theorem*}[The Fourier Inversion Theorem]
Suppose $f$ is integrable and piecewise continuous on $\mathbb{R}$, defined at its points of discontinuity so as to satisfy $f(x)=\frac{1}{2}[f(x-)+f(x+)]$ for all $x$. Then $f(x)=\lim\limits_{\epsilon\rightarrow0}\frac{1}{2\pi}\int e^{-i\xi x}e^{-\epsilon^2\xi^2/2}\hat{f}(\xi)d\xi$, $x\in\mathbb{R}$. Moreover, if $\hat{f}\in L^1$, then $f$ is continuous and $f(x)=\frac{1}{2\pi}\int e^{-i\xi x}\hat{f}(\xi)d\xi$, $x\in\mathbb{R}$.
\end{theorem*}

The uniqueness of the inverse Fourier transform is guaranteed by this theorem. If we can find a function of time, whose Fourier transformation gives back the complex dielectric function $\epsilon^{*}(\omega)$, then this function would be the time derivative of the dielectric relaxation $\epsilon(t)$. We use the following ansatz
\begin{equation}
\frac{e^{-\gamma t}\sin{(Kt)}}{K}H(t)
\end{equation}
where $\gamma=\frac{\nu}{2m}$ and $K=\sqrt{-\frac{\nu^2}{4m^2}+(C\omega_p)^2}$ and $H(t)$ is a Heaviside step function, whose Fourier transformation is expressed as
$\frac{1}{\omega^2-i\nu \omega/m-(C\omega_p)^2}$.

However, we need to put care in taking $\nu \gg 2mC\omega_D$, which amounts to restricting our analysis to the high-friction overdamped dynamical regime. In this way, we finally obtain (for $t>0$)
\begin{equation}
\frac{d\epsilon(t)}{dt}=e^{-\frac{\nu t}{2m}}\int_0^{\omega_D}
\frac{A\rho(\omega_p)\sinh{(\sqrt{\frac{\nu^2}{4m^2}-(C\omega_p)^2}t)}}{\sqrt{\frac{\nu^2}{4m^2}-(C\omega_p)^2}} d\omega_p.
\end{equation}
Upon further integrating over $t$, we recover Eq.(19) in the main article.\\

\section{Behavior of $\epsilon''$ when $\omega\rightarrow0$}
Without loss of generality, we specialize on the Markovian case $\nu=const$ and take the limit $\omega\rightarrow0$ in Eq.(13):
\begin{align}
\lim_{\omega\rightarrow0}\epsilon(\omega)^*&=\lim_{\omega\rightarrow0}\left(1-\int_0^{\omega_D}\frac{A\rho(\omega_p)}{\omega^2-i(\nu/m)\omega-C^2\omega_p^2}d\omega_p\right)\notag\\
&=1-A\lim_{\omega\rightarrow0}\int_0^{\omega_D}\frac{\rho(\omega_p)}{\omega^2-i(\nu/m)\omega-C^2\omega_p^2}d\omega_p.
\end{align}
We Taylor-expand $\rho(\omega_p)$ around $\omega_p=0$:
\begin{equation}
\rho(\omega_p)=\rho(0)+\rho^{\prime}(0)\omega_p+\frac{\rho^{\prime\prime}(0)}{2}\omega_p^2+\frac{\rho^{(3)}(0)}{6}\omega_p^3+...
\end{equation}
Thus, after substituting Eq.(C2) into Eq.(C1), we have
\begin{widetext}
\begin{align}
\lim_{\omega\rightarrow0}\epsilon(\omega)^{*}&=
1-A\lim_{\omega\rightarrow0}\int_0^{\omega_D}\frac{\rho(0)+\rho^{\prime}(0)\omega_p
+\frac{\rho^{\prime\prime}(0)}{2}\omega_p^2+\frac{\rho^{(3)}(0)}{6}\omega_p^3+...}{\omega^2-i(\nu/m)\omega-C^2\omega_p^2}d\omega_p\notag\\
&=1-A\lim_{\omega\rightarrow0}
\int_0^{\omega_D}\frac{[\rho(0)+\rho^{\prime}(0)\omega_p+\frac{\rho^{\prime\prime}(0)}{2}\omega_p^2+\frac{\rho^{(3)}(0)}{6}\omega_p^3+...]
(\omega^2+i(\nu/m)\omega-C^2\omega_p^2)}{\omega^4-2C^2\omega^2\omega_p^2+C^4\omega_p^4+\omega^2\nu^2/m^2}d\omega_p\notag\\
&=1-A\lim_{\omega\rightarrow0}\int_0^{\omega_D}\frac{W_0(\omega)+W_1(\omega)\omega_p+W_2(\omega)\omega_p^2+W_3(\omega)\omega_p^3+...}
{\omega^4-2C^2\omega^2\omega_p^2+C^4\omega_p^4
+\omega^2\nu^2/m^2}d\omega_p
\end{align}
\end{widetext}
where $W_0(\omega)=\rho(0)(\omega^2+i(\nu/m)\omega),W_1(\omega)=\rho(0)^{\prime}(\omega^2+i(\nu/m)\omega),W_2(\omega)=-\rho(0)C^2
+\frac{\rho(0)^{\prime\prime}}{2}(\omega^2+i(\nu/m)\omega),W_3(\omega)=\frac{(\omega^2+i(\nu/m)\omega)\rho^{(3)}(0)}{6}-C^2\rho^{\prime}(0).$
In order to let the integrand be continuous for both real and imaginary part, $(\omega,\omega_p)\in\mathcal{R^{+}}\cup\{0\}\times\mathcal{R^{+}}\cup\{0\}$ (it makes sense to change the order of integration/limit at $(0,0)$), we must have $\rho(\omega_p)\sim0, \rho^{\prime}(\omega_p)\sim0, \rho^{\prime\prime}(\omega_p)\sim0, \rho^{(3)}(\omega_p)\sim0$ as $\omega_p\rightarrow0$. There is no restriction for $\rho^{(4)}(\omega_p)$ or higher order as $\omega_p\sim0$. Hence, we must have $\rho(\omega_p)\sim\omega^n$ for $n>3$. In order to study the scaling of $\epsilon''$ on the left flank of the $\alpha$-peak, we can set $\rho(\omega_p)\sim\omega_p^4$, then from Eq. (C3), we want to study the behavior of
\begin{equation}
\lim_{\omega\rightarrow0}\int_0^{\omega_D}\frac{\omega_p^4}{\omega^2-i\nu/m\omega-C^2\omega_p^2}d\omega_p.
\end{equation}

Without loss of generality for the asymptotic analysis we can take $\omega_{D}=1$, and the integral can be evaluated analytically to the following expression
\begin{widetext}
\begin{align}
    \epsilon''(\omega)&=\int_{0}^{1} \frac{\omega
D(\omega_{p})}{(\omega_{p}^2 - \omega^2)^2 + \omega^2} \, d\omega_{p}\\
    &= \omega + \frac{1}{2} \left( (\omega^2 + i \omega)^{3/2}
\arctan \left[ \frac{i}{\sqrt{\omega^2 + i \omega}} \right] -
(\omega^2 - i\omega)^{3/2} \arctan \left[
\frac{i}{\sqrt{\omega^2 - i \omega}} \right] \right)
\end{align}
\end{widetext}

from which we obtain
\begin{equation}
\epsilon''(\omega) \approx \omega + \frac{\pi}{4} ((\omega^2 +i \omega)^{3/2} - (\omega^2 - i\omega)^{3/2})
\end{equation}
and hence $\epsilon''(\omega)\sim\omega$ in the limit of small frequency, in agreement with the experimental data.

\end{appendix}


\begin{thebibliography}{99}
\bibitem{note}
The response in the time domain is related to the response in the frequency
domain~\cite{Montroll}
\begin{align}
\epsilon^*(\omega)&=\epsilon'(\omega)-i\epsilon''(\omega) \notag \\
&=\epsilon^*(\infty)+(\epsilon^*(\infty)-\epsilon^*(0))\int_{0}^{\infty}
e^{-i\omega t}\frac{d\phi}{dt} dt.
\end{align}
The relaxation function is related to the response in the time domain
via~\cite{Montroll}:
\begin{equation}
\begin{split}
\epsilon(t)&=\epsilon^*(\omega=\infty)+{}\\
&(\epsilon^*(\omega=0)-\epsilon^*(\omega=\infty))(1-\phi(t))
\end{split}
\end{equation}
where $\phi(t)=\exp{[-(t/\tau)^{\beta}]}$.
Predictions of this equation have also been plotted for comparison in Fig. 2-3, for the frequency-dependent response, and in Fig. 4 for the time-dependent response, together with the predictions of our theory.
Note that the parameters in the Kohlrausch model have been also adjusted to
match with the experimental data.
\end{thebibliography}
\end{document}